\documentclass[useAMS,usenatbib]{mn2e} 
\usepackage{aas_macros}
\usepackage{graphics}
\usepackage[pdftex]{graphicx}
\usepackage{epstopdf}
\usepackage{epsfig}  
\usepackage{natbib} 
\usepackage{float}
\usepackage{amsmath}
\usepackage{times}
\usepackage[varg]{txfonts}
\usepackage{verbatim} 
\usepackage{multirow,bigdelim} 
\usepackage{color}

\usepackage[normalem]{ulem}

\def  \LCDM{$\Lambda$CDM}

\def \kms {{\rm km s$^{-1}$}}

\newcommand{\hmpc}{{\,\rm h^{-1}Mpc}}
\def\br{{\bf r}}

\def\bv{{\bf v}}
\def\Vbulk{{\bf V}$_{\rm bulk}$}
\def\VbulkR{{\bf V}_{\rm bulk}(R)}
\def\uo{u^o}
\def \WF {^{\rm WF}}
\def \CR {^{\rm CR}}

\title[CF2 Bulk Flow]{Cosmic Bulk Flow and the Local Motion from Cosmicflows-2}

\author[Hoffman et al.]
{Yehuda Hoffman$^1$,
H\'el\`ene M. Courtois$^{2}$ and R. Brent Tully$^3$\\
$^1$Racah Institute of Physics, Hebrew University, Jerusalem 91904, Israel\\
$^2$University of Lyon; UCB Lyon 1/CNRS/IN2P3; IPN Lyon, France\\
$^3$Institute for Astronomy (IFA), University of Hawaii, 2680 Woodlawn Drive, HI 96822, USA\\
}

\begin{document}
\date{Submitted to MNRAS August 4th, 2014}

\pagerange{\pageref{firstpage}--\pageref{lastpage}} \pubyear{2014}

\maketitle

\label{firstpage}

\begin{abstract}

Full sky surveys of peculiar velocity are arguably the best way to map the large scale structure (LSS) out to distances of a few$\times 100\hmpc$. 
Using the largest and most accurate ever catalog of galaxy peculiar velocities {\it Cosmicflows-2}, the LSS has been reconstructed by means of the Wiener filter (WF) and constrained realizations (CRs) assuming as a Bayesian prior model the 
\LCDM\ model with the WMAP inferred cosmological parameters.
The present paper focuses on studying the bulk flow of the local flow field, defined as the mean velocity of  top-hat spheres with radii ranging out to $R=500\hmpc$.
The estimated LSS, in general, and the bulk flow, in particular, are  determined by the tension between the observational data and the assumed prior model. A prerequisite  for such an analysis is the requirement that the estimated bulk flow is consistent with the prior model. Such a consistency is found here.
At $R=50\ (150) \hmpc$ the estimated bulk velocity is $250 \pm 21 \ \  (239 \pm 38)$~\kms. The corresponding cosmic variance at these  radii  is $126$ ($60$)~\kms, which implies that these estimated bulk flows are dominated by the data and not by the assumed prior model.  The estimated bulk velocity is dominated by the data out to $R\approx 200 \hmpc$, where the cosmic variance on the individual Supergalactic Cartesian components (of the r.m.s. values) exceeds the variance  of the CRs by at least a factor of 2. The SGX and SGY components of the CMB dipole velocity are recovered by the WF velocity field down to a very few \kms. The SGZ component of the estimated velocity, the one that is most affected by the Zone of Avoidance, is off by $126$\kms (an almost 2 sigma discrepancy).
The bulk velocity analysis  reported here is virtually unaffected by the Malmquist bias and very similar results are obtained for the data with and without the bias correction.

\end{abstract}

\begin{keywords}
(cosmology:) large-scale structure of universe
\end{keywords}

\section{Introduction}
\label{sec:intro}

In the standard model of cosmology the large scale structure (LSS) of the universe grows out of a primordial perturbation field via gravitational instability. 
The continuity equation implies that the evolving density field is associated with a peculiar velocity field, both of which represents departures,  or fluctuations, from a pure Hubble expansion. 

The Cosmic Microwave Background (CMB) dipole anisotropy, which is interpreted as the consequence of the peculiar motion of the Local group with respect to the CMB is the best evidence and example of that motion. First hints for the dipole anisotropy were given by \cite{1969Natur.222..971C} and  \cite{1971Natur.231..516H} and a more definitive determination by \cite{1977PhRvL..39..898S}. These discussions have set the stage for an effort that continues to the present epoch to map the three dimensional (3D) velocity field in our 'local universe', namely our local cosmic neighbourhood \citep{1981ApJ...246..680T,
1982ApJ...258...64A,
1988ApJ...326...19L,
1995ApJ...454...15S,
1997ApJS..109..333W,
2007ApJS..172..599S,
 2008ApJ...676..184T}.
The motivation for these studies is twofold. One is the wish to map the underlying mass distribution in the local universe in general and the quest for the sources that induce the CMB dipole in particular. The other is to use the recovered local 3D flow field as a probe on the statistics of the underlying primordial perturbation field. This motivation leads immediately to the use of observed velocities as probes of the values of the cosmological parameters.

The present epoch 3D velocity field is affected by non-linear processes, occurring mostly on small spatial scales. 
This has led people to focus on bulk velocities, namely the mean velocity of a large volume of space, as a mean of filtering out small scale non-linear motion and thus concentrate on the large scale modes that  affect the bulk velocity \citep[e.g.][]{1980lssu.book.....P}. The common choice is the definition  of bulk velocity as the mean of  a spherical top-hat window function convolved with the 3D velocity field. However, there are exception to this practice. The amplitude, scale dependence, rate of convergence and the direction of the local bulk velocity are the subject of many studies 
\citep[summary in ][]{2009MNRAS.392..743W}.

Recent peculiar velocity studies have found contradictory results in the amplitude of the
cosmic bulk flow on scales of $ 50-100 \hmpc$ in excess of that expected in the $\Lambda$ cold dark matter (\LCDM) model  \citep{2009MNRAS.392..743W,2010ApJ...709..483L,2010MNRAS.407.2328F,2011MNRAS.414..264C} finding higher, seemingly inconsistent with \LCDM,  values
while \citep{2011ApJ...736...93N,2012MNRAS.420..447T,2014MNRAS.437.1996M}  finding lower values.
Current surveys of peculiar velocities extend out to roughly 100$\hmpc$, e.g. the SFI++ survey \citep{2006ApJ...653..861M,2007AAS...211.5505S},
and consequently current estimation of the bulk velocity do not go beyond that scale. Attempts to determine the bulk flow on larger scales are often based on using the anisotropic pattern of large-scale galaxy clustering, known as the redshift-space distortion \citep{kaiser}. A novel method based on the apparent dimming or brightening due to peculiar motion of galaxies in a redshift survey has been recently suggested \citep{2011ApJ...735...77N}.


Surveys of peculiar velocities are notorious for being  'difficult data to use'. 
Observed radial velocities are noisy.
The noise is driven by the uncertainties in the distance estimations.   For example, the intrinsic scatter in the Tully-Fisher (TF) relation \citep{1977A&A....54..661T} leads to a $\sim$20\% relative error in the distance, hence also in the estimated velocities. The relative error depends only weakly on the distance.
 Peculiar velocities datasets are sparse - the CF2 catalog consists of roughly 8000 data points within roughly a $100 \hmpc$. The Galactic zone of avoidance (ZOA) limits the sky coverage of velocities surveys. 
The analysis of observational peculiar velocities datasets is hampered by various types of Malmquist bias. In the case of cosmic-flows2, the major one is due to the large errors on individual distance determinations.
 Such "difficult data to use" is best handled within the Bayesian framework of the Wiener filter (WF) and constrained realizations (CRs) of Gaussian fields \citep{1991ApJ...380L...5H,1995ApJ...449..446Z,1999ApJ...520..413Z}.

We are using the largest ever and offered as the most accurate catalog of galaxy peculiar velocities, the Cosmicflows-2 dataset (CF2) \citep{2013AJ....146...86T}. The value of the Hubble constant used in the survey required for the decomposition of cosmic expansion and peculiar velocity components is calibrated by Type Ia supernovae  calibration at $z \sim 0.1$ \citep{2012ApJ...758L..12S}.
The full 3D velocity field estimated from the CF2 dataset by means of the WF/CRs methodology has been recently presented in \cite{Tully-Nature2014}. The main aim of the present paper is to describe the flow field around us by means of its bulk velocity, namely the mean velocity obtained by convolving the full 3D velocity field with a given window function. Bulk velocities constitute a very compressed representation of the full 3D velocity fields and their analysis has played a leading role in the study of the cosmic flow field. We shall follow that tradition and dedicate the majority of the paper to the estimation of bulk velocity of the local universe. A secondary goal of the paper is the study of the motion of the Local Group (LG) with respect to the CMB. The problem to be addressed here is the assessment  of how well the estimated LG velocity agrees with the observed CMB dipole.

The structure of the paper is as follows. 
The CF2 data is described in \S \ref{sec:data}.
The WF/CRs methodology and its application to the CF2 data are reviewed in \S \ref{sec:method}.
\S \ref{sec:results} presents a comprehensive review of the results. 
\S \ref{sec:disc} concludes the paper with a summary and a final discussion, 

\section{Data : CosmicFlows-2}
\label{sec:data}

Cosmicflows-2 is the second generation catalog of galaxy distances
built by the Cosmicflows collaboration. Published in \cite{2013AJ....146...86T}, it contains more than 8,000 galaxy peculiar velocities.
Distance measurements come mostly from the Tully-Fisher relation
and the Fundamental Plane methods (\cite{2001MNRAS.321..277C}).  Cepheids (\cite{2001ApJ...553...47F}), Tip of the Red Giant
Branch (\cite{1993ApJ...417..553L}), Surface Brightness Fluctuation (\cite{2001ApJ...546..681T}), supernovae of type Ia (\cite{2007ApJ...659..122J}) and other miscellaneous methods also contribute to this large dataset but to a minor extent (12\%). 

For cosmic bulk flow purpose, in order to work above the scale of non-linearity in galaxy cosmological displacements
(virial motions in clusters, in knots of filaments, galaxy pairing interactions, etc..) we use in this paper the grouped version of 
cosmicflows-2 containing 4,838 peculiar velocity values. Grouping also adds value to the following analysis since it decrease 
the statistical errors compared to individual galaxy values ($\sim$ 20\%). The mean error for the groups is $\sim$ 9\% of the distance.
See Tully et al. (2013) for details of the distribution of error estimates for components of the grouped catalog.

\section{Method}
\label{sec:method}

\subsection{Bayesian methodology}
The WF/CRs methodology has been extensively used to reconstruct the underlying density and three dimensional velocity fields from data bases of  galaxy distances and peculiar velocities, starting with \cite{1993ApJ...415L...5G}. A full Bayesian framework of the WF/CRs methodology was developed in  \cite{1995ApJ...449..446Z} and its first application to peculiar velocities data base was presented by Zaroubi, Hoffman \& Dekel (1999).
The present WF/CRs reconstruction of the LSS follows the formalism of \cite{1999ApJ...520..413Z}. The methodology was applied to the SEcat catalog of peculiar velocities (\cite{2002MNRAS.331..901Z}) and more recently to the Cosmicflows-1 data base \citep{2012ApJ...744...43C,Courtois2013} and the Cosmicflows-2 data base \citep{Tully-Nature2014}. 

 As mentioned earlier, the WF/CRs methodology recovers the linear density and 3D velocity fields. This is certainly a bad assumption for the density field but a good one for the velocity field in the quasi-linear regime, namely away from virial motions in groups and clusters of galaxies. As for the linear density field, the linear WF/CRs essentially recovers the divergence of the velocity field assuming the linear theory relation between the two fields. As the current paper focuses on the bulk velocity on scales of a few Megaparsecs and above the linear theory provides the adequate theoretical framework within which all present calculations are conducted.

\subsection{Prior model}
The standard model of cosmology, a flat $\Lambda$ cold dark matter (\LCDM) cosmology, is taken here as the prior model. The Wilkinson Microwave
Anisotropy Probe 5 (WMAP5) cosmological parameters are assumed (\cite{2009ApJS..180..330K}). 
The following parameters, in particular, have been used here:  
$\Omega_m=0.28$ (the mass density parameter), $h=0.70$ (the Hubble constant in units of $100$\kms Mpc$^{-1}$) and $\sigma_8=0.817$ (the r.m.s. of the linear density fluctuations in a sphere of $8 \hmpc$). 
The results reported here remain virtually unchanged for the later release of cosmological parameters of the WMAP team.  
The model further assumes that the primordial fluctuations constitute a Gaussian random field.
 
\subsection{Computational details}
Three kinds of velocities field are considered here: 1. A field obtained by applying the WF to the Cosmicflows-2 data (${\bf V}^{\rm WF}$); 
2.  A  realization constrained by the Cosmicflows-2 data (${\bf V}^{\rm CR}$); 
3. An unconstrained, i.e.  random, realization (${\bf V}^{\rm RAN}$). The construction of a CR involves a one-to-one mapping from a random to a constrained realization and all the random realizations used here are the ones used to construct the CRs. 
The random realizations are constructed by using the Fast Fourier Transform (FFT) technique. Hence they obey periodic boundary conditions.

The different fields are constructed on a Supergalactic Cartesian $256^3$ grid centered on the Milky Way galaxy. Three different computational boxes are used here with side length of $L=320,\ 1280$ and $2560 \hmpc$ (referred to as BOX320, BOX1280, etc.)
The WF field has been constructed on the BOX320. The other larger boxes are used for the construction of the constrained and random fields. These large boxes are used to perform convergence tests on the bulk velocity of the CR fields.

 A note is due here on the role of the periodic boundary conditions and the construction of the WF, CR and RAN fields. The evaluation of the WF field at a given point in space relies on the calculation of various 2-point correlation  functions. The analytical form of these functions is given by one dimensional (1D) semi-infinite integrals over the power spectrum convolved with functions such as the spherical Bessel functions \citep{1999ApJ...520..413Z}. The integrals are evaluated by standard techniques of numerical 1D integrals. The value of the field at a given point is unaffected by the size of the computational box as the integrals used to calculate the correlation functions  assumes an infinite space. The random realizations, on the other hand, obey periodic boundary conditions and therefore are devoid of any possible power on scales larger than the scale of the box.  As velocities are induced by long waves the size of the computational box reflects the tension between the needed resolution (i.e. small scales) and the need to capture the power coming from long waves. The CRs are composed of the sum of the WF mean field and the random component and as such they are partially affected by the periodic boundary conditions. The large computational boxes used here are chosen to minimize the effect and the convergence tests confirm it.

\subsection{Malmquist bias }

 The study of issues associated with various biases that can arise in the measurement of distances to astronomical objects has a long history, dating from the pioneering studies of Gunnar Malmquist \citep{Malmquist1920, Malmquist1922, Malmquist1924}.

In CF2 \citep{2013AJ....146...86T} steps were taken to provide distances free of those that are often called the selection bias \citep{1994ApJS...92....1W}, 
but also confusingly referred to as simply the Malmquist bias.
Even with statistically unbiased distances, however, the forms of Malmquist bias due to scattering by errors from regions of greater sampling to regions of lesser sampling are expected (whether of the so-called `homogeneous' or `inhomogeneous' varieties).  Also one must contend with the asymmetry in amplitude of peculiar velocity errors that arises from the translation of uncertainties that are symmetric in the logarithm of distances.  In the following, it is these latter biases that will be addressed with reference to Malmquist biases.

The WF reconstructed velocity field is used here to correct for the Malmquist bias. A full description and testing of the procedure is to be published elsewhere and only a short description is presented here. 
The procedure is done in  two steps. In the first the WF is applied to the raw data. New estimated distances are  defined by  subtracting from the observed 
radial velocity of a data point the radial component of the WF velocity field at its position. This adjustment brings the estimated distances to be much closer to their position in redshift space. The Wiener Filter is then applied to the revised catalog to yield the reconstructed flow that is presented here. The Wiener Filter correction of the Malmquist bias has been tested against mock catalogs drawn from N-body simulations constrained by the CF2 data to reproduce the actual universe. 

The Malmquist bias introduces a spurious strong monopole term into the reconstructed velocity field but is expected to hardly affect the bulk velocity which is associated with the dipole of the velocity field. Nevertheless all the analysis presented here is performed on the the Malmquist bias corrected data. A comparison of the bulk velocity determined from the corrected and uncorrected data is 
presented in Appendix  \ref{app-MBc}.

We have not attempted here to perform a maximum likelihood analysis to find the most likely \LCDM\ model, but rather decided to choose an a priori  WMAP5 set of cosmological parameters. This procedure leaves us with only one free parameter that enters the data covariance matrix, $\sigma_\ast$, which serves as a crude proxy to small scales non-linear motion. We fine tune  its value by the requirement that it makes the $\chi^2/$d.o.f. equal to unity. Indeed for the CF2 data a $\sigma_\ast=100$~\kms\ we find a $\chi^2/{\rm d.o.f. }=1.002$ for the raw data and $1.018$ after applying the WF Malmquist bias correction. This small value of $\sigma_\ast$ is appealing as the data consists of a grouped version of the CF2 data base which 
largely negates non-linear motions associated with groups and otherwise 
contains mostly spiral galaxies for which non-linear motions are quite insignificant. A comparison can be made with the $\sigma_\ast=250$\kms\  adopted by \cite{2001MNRAS.326..375Z} in their study of a catalog of nearby early-type galaxies. In any case such a small $\sigma_\ast$ is relevant only for the determination of the very  local velocities and it does not affect  the bulk flow analysis. 

\subsection{Bulk velocity determination}
Given the full three dimensional (3D) velocity field, evaluated on a Cartesian grid, the bulk velocity vector is readily calculated as the volume weighted mean velocity field within a top-hat sphere of radius $R$. The  top-hat term is used here to denote a sphere with sharp cutoff. The bulk velocity is defined by:
\begin{equation}
{\bf V}{^{\rm (T)}_{\rm bulk}}(R) = {3\over 4 \pi R^3} \int_{r < R} {\bf v}^{\rm (T)}({\bf r}) \ {\rm d}^3 r
\label{eq-Vbulk}
\end{equation}
(Here, T stands for the type of the field - WF, CR or RAN.)

Given that  $\VbulkR$ is obtained by a linear convolution over the full 3D velocity field, it follows that to the extent that $ {\bf v}({\bf r}) $ is a Gaussian random field then $\VbulkR$  is a Gaussian random vector. In the absence of constraining data, the prior expectation of 
of the bulk velocity is the null vector. For a large enough ensemble of random realizations the mean of  $\VbulkR$ over the ensemble is zero, $ \Big< {\bf V}{^{\rm (RAN)}_{\rm bulk}}(R) \Big> = 0$. The mean field of the ensemble of  CRs is the WF field, hence 
$ \Big< {\bf V}{^{\rm (CR)}_{\rm bulk}}(R) \Big> = {\bf V}{^{\rm (WF)}_{\rm bulk}}(R)$. Given the finite number of realizations used here we expect some statistical deviations from the mean value. The variance of the dispersion of the ensemble of CRs and RANs from their 
mean values is used here as a proxy to the theoretical variance. In the following we use the term 'sigma' as a shortcut for the standard deviation of the scatter around the mean. The statistical significance of the results will be measured by units of 'sigma'.

\section{Results}
\label{sec:results}

\begin{figure*}
\includegraphics[width=0.55\textwidth]{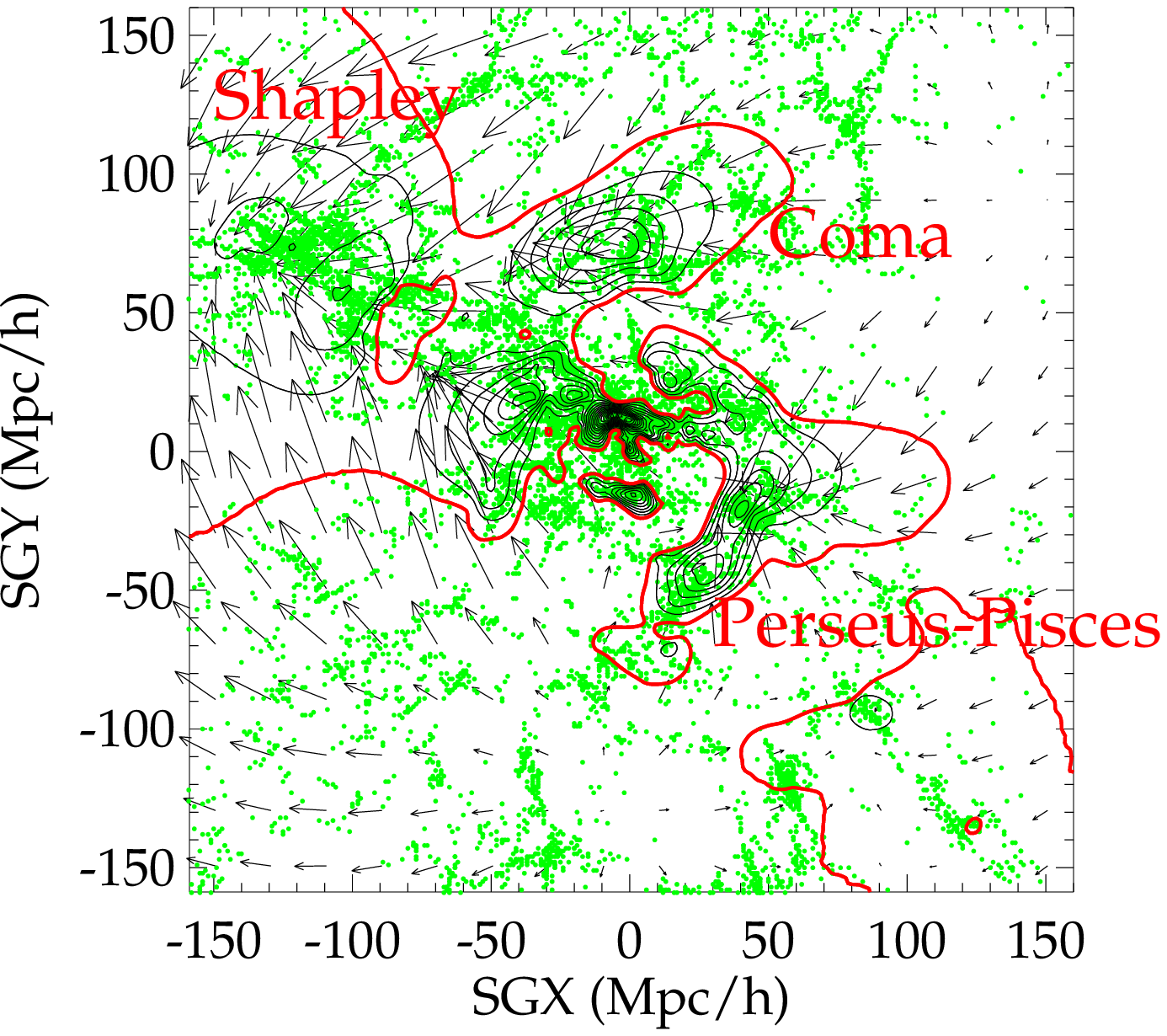}
\hspace{-1cm}
\includegraphics[width=0.49\textwidth]{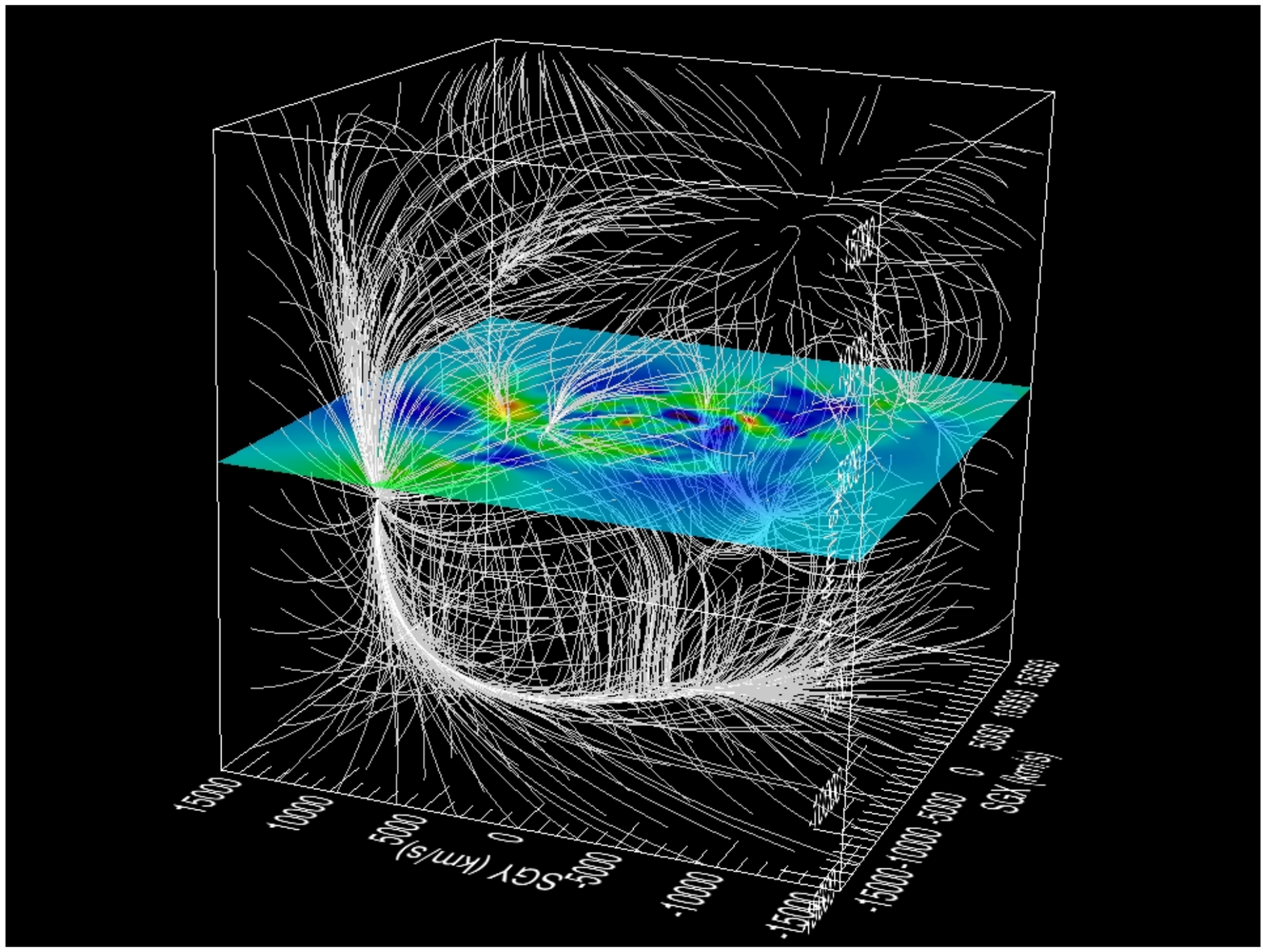}
\caption{
Left panel: The WF 3D velocity field (represented by vector arrows) and the linear density field on the Supergalactic Plane. Red contour denotes the zero over-density and the black contours the positive over-densities. Galaxies from the 2MASS redshift survey are presented by green dots and are placed in redshift space.
Right panel: 
The WF of the 3D velocity field (represented by flow lines) in a box of $L=320\hmpc$ centered on the LG. The density field on the Supergalactic Plane is presented as well.}
\label{fig:overview}
\end{figure*}

We start by presenting an overview of the  WF reconstructed density and 3D velocity field. Figure \ref{fig:overview} presents the WF density and velocity field at the equatorial plane in supergalactic coordinates and the velocity field in the BOX320 in a 3D presentation. The Supergalactic Plane cut also contains the redshift space distribution of galaxies in an update of the 2MASS Extended Source Catalog \citep{2000AJ....120..298J}. The 3D presentation of the flow field shows the dominant role of the Shapley supercluster in shaping the local flow.

\subsection{Bulk velocity}

\begin{figure*}
\includegraphics[width=0.75\textwidth]{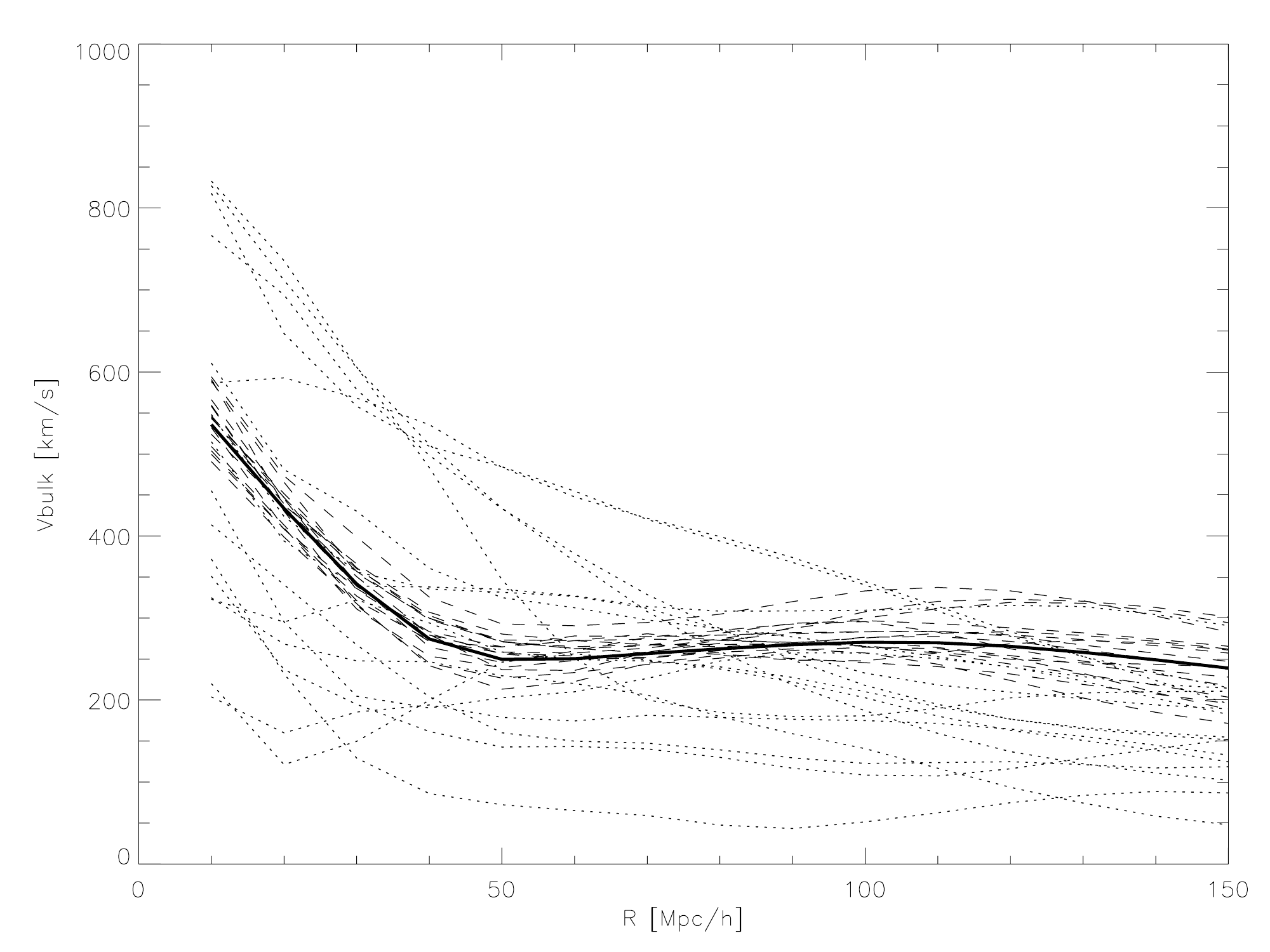}
\includegraphics[width=0.75\textwidth]{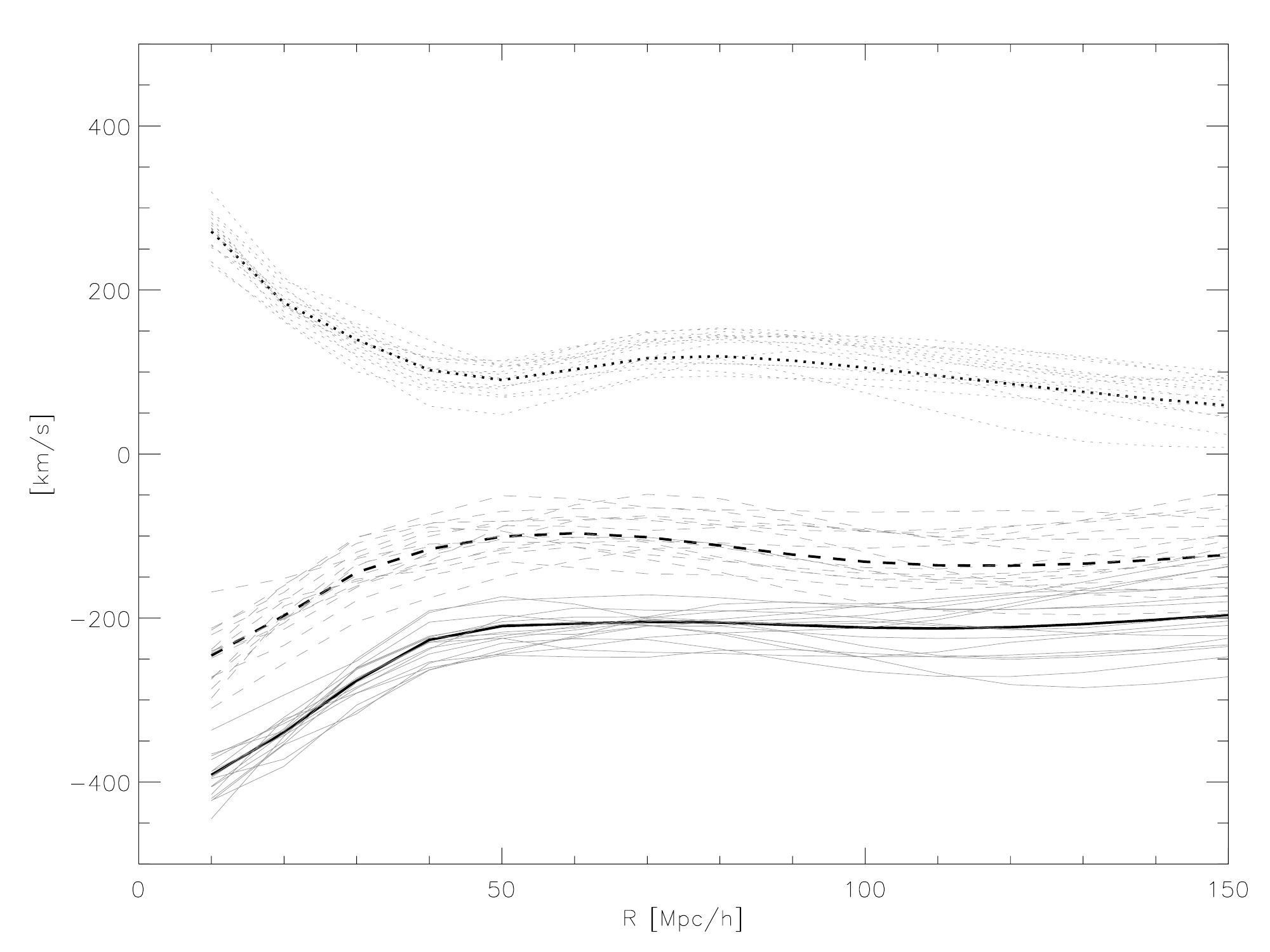}
\caption{
Upper panel: Bulk flow of a top-hat sphere as a function of its radius of: the Wiener Filter reconstructed velocity filed (thick full line); an ensemble of 16 constrained realizations  (thin dashed lines) and of an ensemble of random realizations (thin dotted lines). 
Lower panel: 
The three (Supergalactic) Cartesian components (SGX (full lines), SGY (dotted lines) and SGZ (dashed lines) ) of an ensemble of 16 constrained realizations (BOX1280, thin lines). The thick lines represent the WF.
} 
\label{fig:Vbulk}
\end{figure*}

Figure \ref{fig:Vbulk} presents the main results of the paper. It shows the bulk velocity of the WF, 16 (BOX1280) constrained realizations and the matching unconstrained ones 
as a function of depth over the range of $R=10$ to $150\hmpc$. The compatibility of the estimated WF bulk velocity with the predictions of the \LCDM/WMAP5 model is clearly manifested. The amplitude of WF bulk velocity lies within the expected scatter over the range of  $R=[10 - 150]\hmpc$. At $R=150\hmpc$ the estimated amplitude is $V_{\rm bulk}= (239 \pm 38)$ \kms. It follows that we detect a bulk flow  at more than a 6 sigma significance at roughly the edge of the data. Some representative values as a function of $R$ are given in Table \ref{table:Vbulk}.
The lower panel of Figure~\ref{fig:Vbulk} shows the radial dependence of the three Supergalactic Cartesian components of the estimated WF,  ${\bf V}{^{\rm (WF)}_{\rm bulk}}(R)$. The plots also show also the Cartesian components of an ensemble of 16  ${\bf V}{^{\rm (CR)}_{\rm bulk}}(R)$ (BOX160). At $R=150\hmpc$ one finds:
$V_{\rm bulk \  [SGX,\ SGY,\ SGZ]} = [-196 \pm 43, 59 \pm 27, -123 \pm 41 ]$\kms
The flow is mostly within the Zone of Avoidance (ZOA) which almost coincides with the SGX-SGZ plane at SGY=0. The smallest component of the bulk flow is in the SGY direction. The largest component is in the negative SGX direction, detected with a 4.5 sigma significance.

\begin{table}
\begin{tabular}{lllll}
\hline
R [$\hmpc$]  &  WF  &  CRs  & RANs    \\
\hline
 10     &        536     &       543   $\pm$    33      &       511    $\pm$    214    \\
 40     &       274      &       282    $\pm$   23      &       322    $\pm$    147    \\
 50     &       250      &       258    $\pm$   21      &       292    $\pm$    126    \\
100    &        270     &       279    $\pm$  23      &       209    $\pm$     82    \\
150    &        239     &       241    $\pm$  38      &       161    $\pm$     60    \\
\hline
\end{tabular}
\caption{The bulk flow (in \kms) of spheres of radius $R$ of the Wiener Filter reconstructed field and for an ensemble of 16 constrained (CRs) and random (RANs) realizations.  For mean and standard deviations are presented for the two ensembles.
}
\label{table:Vbulk}
\end{table}

\begin{figure*}
\includegraphics[width=0.75\textwidth]{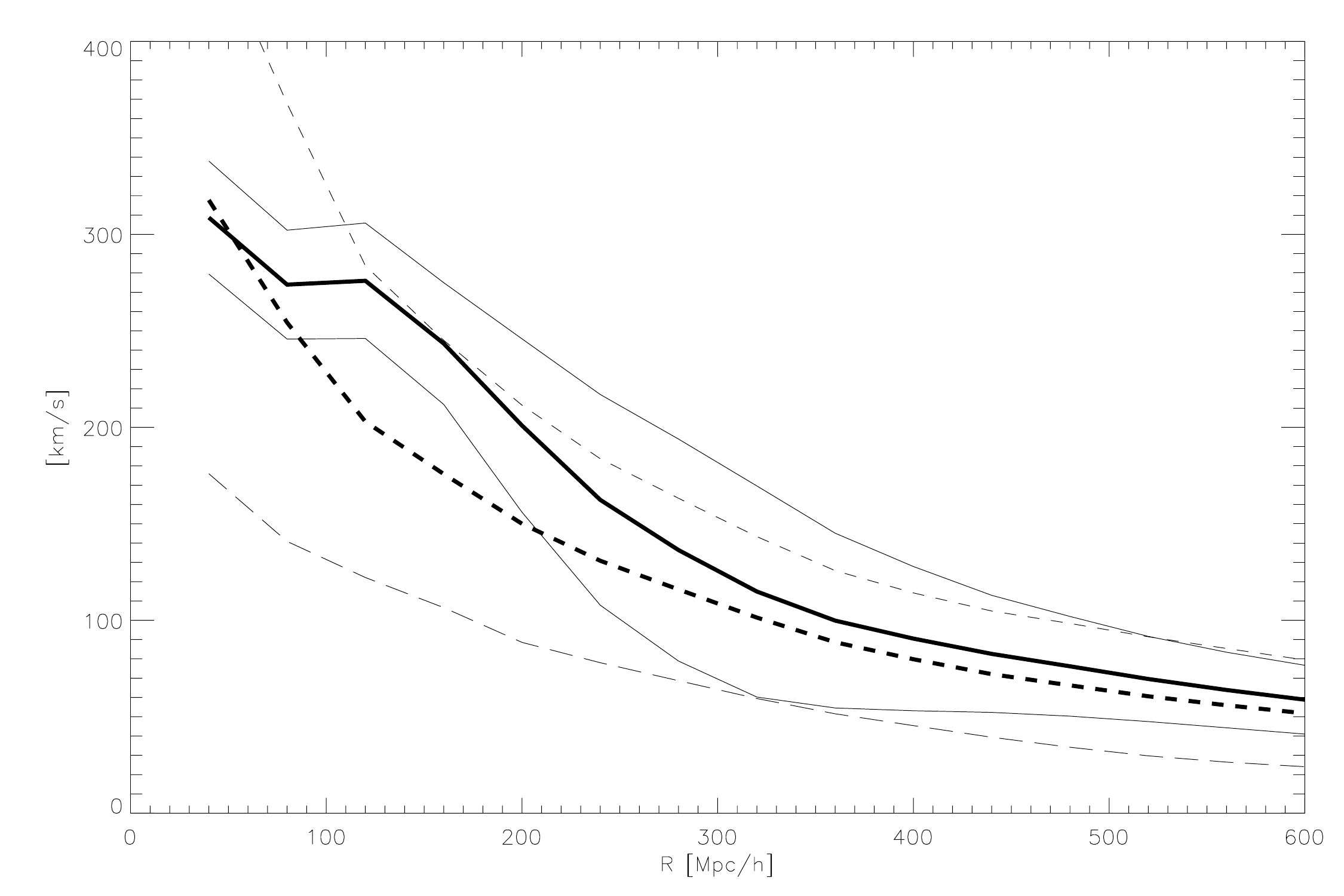}
\includegraphics[width=0.75\textwidth]{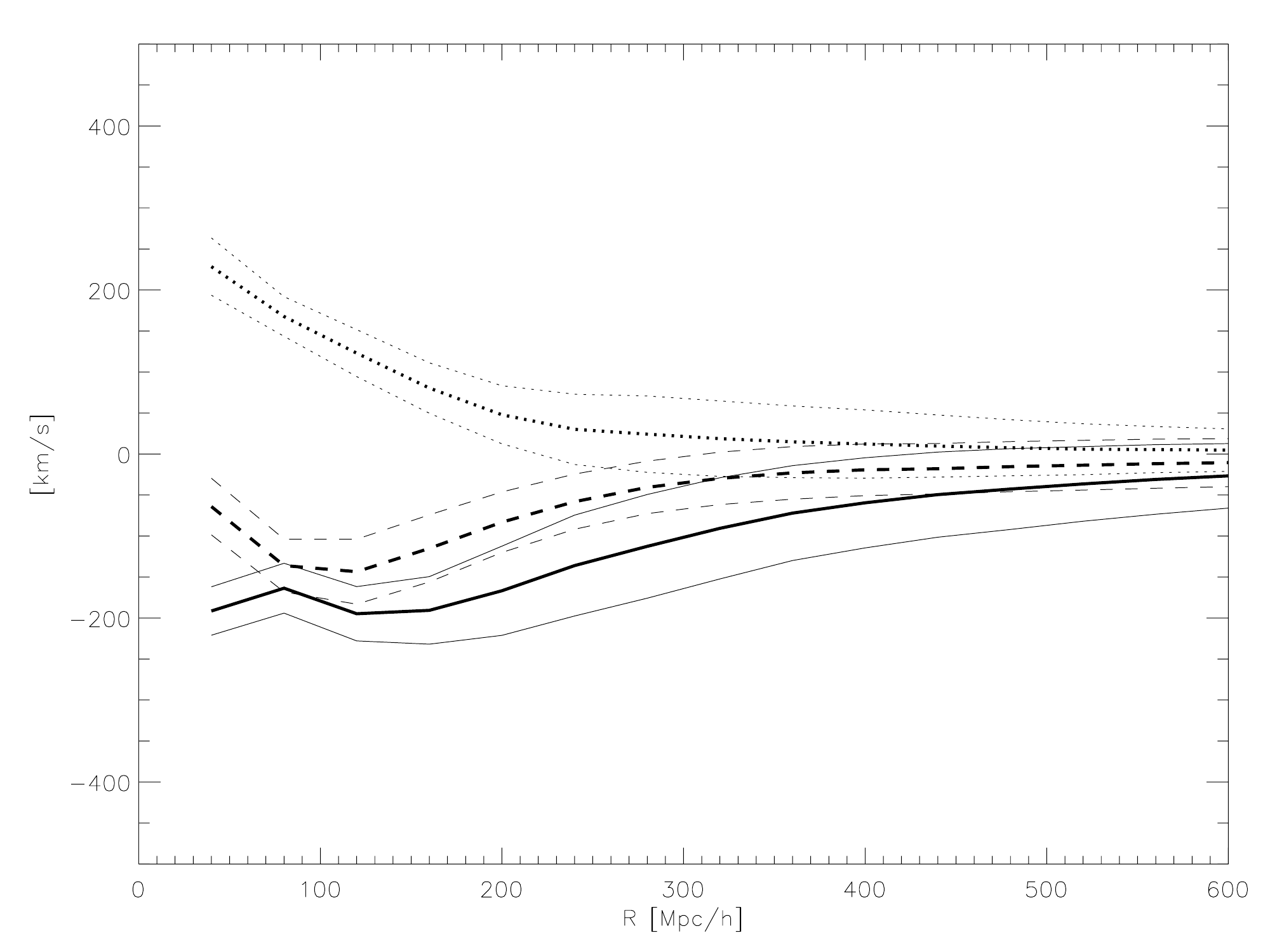}
\caption{
Upper panel: Amplitude of the bulk velocity: The mean (thick lines) and the mean $\pm$ one standard deviation (thin lines) of an ensemble of 20 constrained (solid lines) and unconstrained (dashed lines) realizations.
Lower panel:  Mean (thick lines)  and the mean $\pm$ one standard deviation (thin lines) of the three Cartesian components of the bulk velocity taken over the ensemble of 20 CRs (BOX2560) over the range  $R=[0 - 500]\hmpc$: 
SGX (full lines), SGY (dotted lines) and SGZ (dashed lines) 
}
\label{fig:Vbulk-big}
\end{figure*}

Using the largest computational box, BOX2560, the mean and the scatter around the mean of the amplitude of the bulk velocity of an ensemble of 20 CRs and their corresponding RANs are shown in Figure \ref{fig:Vbulk-big} all the way up to $R=600\hmpc$. Some representative values are presented in Table \ref{table:Vbulk-BOX2560}.
Again, the mean bulk velocity lies within the 1 sigma bracket of the assumed prior model. At $R=600\hmpc$ the WF yields $V_{\rm bulk}= (59 \pm    18)$ \kms, a 3.2 sigma detection. The expected amplitude for random realizations is $ (52 \pm    28)$ \kms. The comparison of the individual Cartesian components is even more illuminating. Considering the limited number  of realizations employed here we expect to find statistical scatter in the values of mean quantities. For random realizations one expects the mean of the Cartesian components of the bulk velocity to be zero and the scatter to be isotropic. The deviation from these expectations enables one to estimate the statistical uncertainty in our results. At $R=600\hmpc$ the mean over the ensemble of RANs of the individual components is a very few \kms. We find here a non vanishing, at the 1 sigma level, SGX/SGY/SGZ component of the bulk velocity out to $R\approx 400/200/300\ \hmpc$.

\begin{table*}
\begin{tabular}{c c  c c c c c c c c c}
\hline\hline
      & \multicolumn{4}{c}{CRs} & \multicolumn{6}{c}{\ \ \ RANs}\\
\hline
   R          &   & $V_{\rm bulk}$          &      $V_{\rm x}$        &           $V_{\rm y}$         &         $V_{\rm z}$              &  &         $V_{\rm bulk}$              &                $V_{\rm x}$         &            $V_{\rm y}$           &         $V_{\rm z}$  \\
   \hline\hline
   40     & \ \ \   &      309    $\pm$   29     &    -191     $\pm$    30   &    229     $\pm$    35   &    -64      $\pm$   34    & \ \ \    &   318      $\pm$   142    &      -12      $\pm$   178    &    19     $\pm$    212    &    40    $\pm$     218  \\
  120    &   &     276      $\pm$    30      &   -195    $\pm$     33     &  123    $\pm$     29    &  -144    $\pm$     40    &   &   254     $\pm$    113     &      -6    $\pm$     127    &    30     $\pm$    138   &     20     $\pm$    116 \\
  200    &   &     201      $\pm$    45      &   -167     $\pm$    54    &    48     $\pm$    35     &  -83    $\pm$     37      &   &   150     $\pm$     61       &     -9     $\pm$     97      &  17     $\pm$    103     &   17    $\pm$      82  \\
  320    &   &      115     $\pm$     55     &     -91    $\pm$     62    &    19    $\pm$     46    &   -30    $\pm$     32       &   &     101   $\pm$       42     &     -1    $\pm$      71     &    9     $\pm$     70     &   15     $\pm$     47  \\
  400    &   &     90     $\pm$     37      &    -60    $\pm$     55    &    12     $\pm$    42    &   -19    $\pm$     31        &   &    80     $\pm$     34       &    4     $\pm$     61        &  6      $\pm$    53      &  10     $\pm$     34  \\
  520    &   &    70     $\pm$     22       &    -36     $\pm$    45    &     6    $\pm$     31    &   -14    $\pm$     30       &   &     61     $\pm$     31       &     5    $\pm$      50       &    2     $\pm$     38    &     5    $\pm$      28  \\
  600    &   &     59     $\pm$     18      &     -27     $\pm$    39    &     5    $\pm$     26     &  -11    $\pm$     29          &   &     52    $\pm$      28      &    6    $\pm$      43      &   2      $\pm$    31       &  4     $\pm$     27  \\
   \hline\hline
\end{tabular}
\caption{The mean and standard deviation of the amplitude and the SGX, SGY and SGZ components of 20 CRs (columns 2-5) and their corresponding RANs (columns 6-9). (CRs and RANs are calculated within BOX2560.)
}
\label{table:Vbulk-BOX2560}
\end{table*}

\subsection{Alignment}

\begin{figure*}
\includegraphics[width=0.75\textwidth]{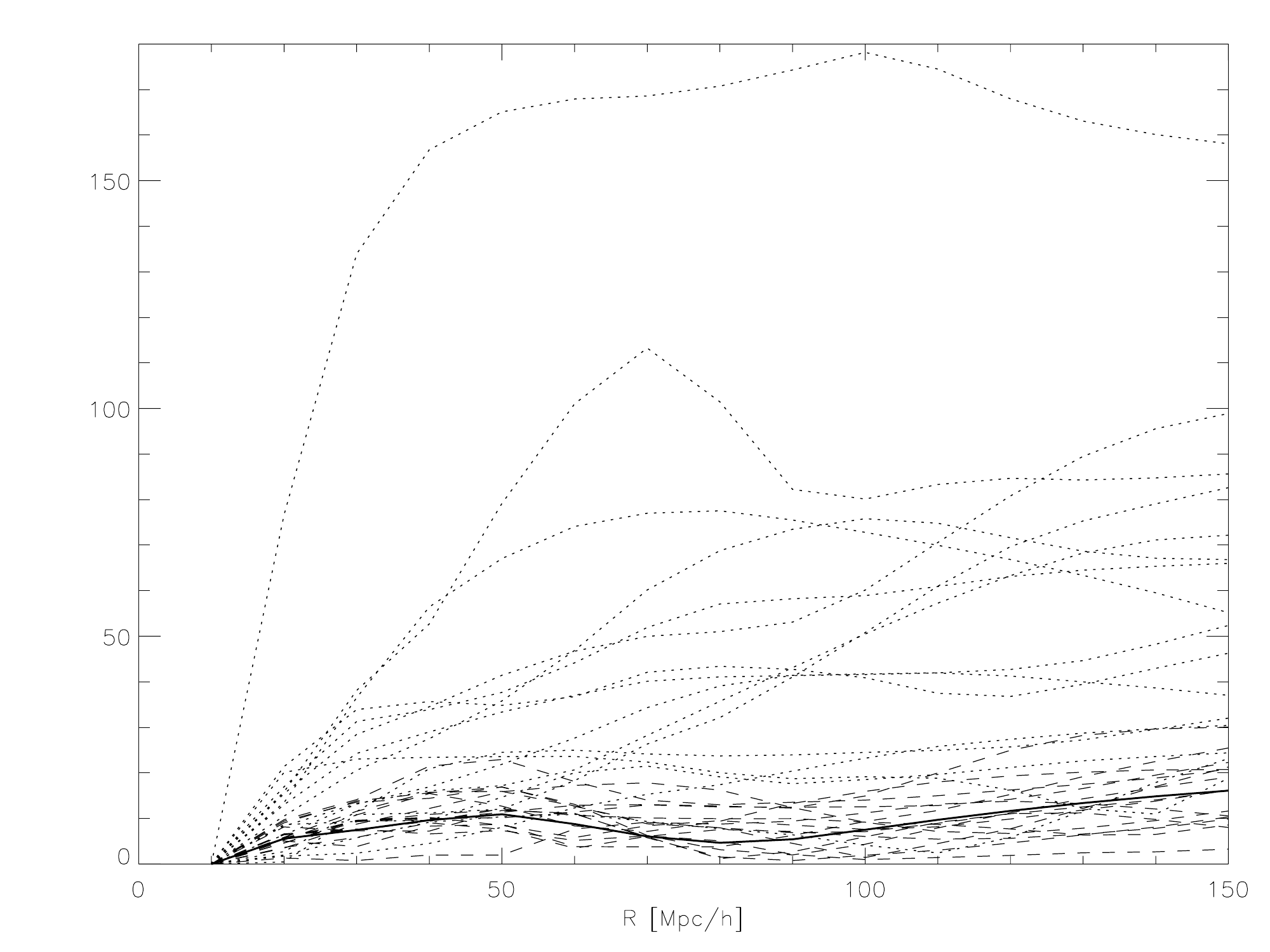}
\caption{Alignment of the bulk velocity with itself as a function of depth, namely of ${\bf V}_{\rm bulk}(R)$ with $ {\bf V}_{\rm bulk}(R=10\hmpc)$, given in degrees: the WF (thick full line), an ensemble of 16 constrained realizations  (dashed lines) and an ensemble of 16 random realizations  (dotted lines).
}
\label{fig:align}
\end{figure*}

The direction of \Vbulk\ remains coherent over a large range of scales. Is it consistent with the \LCDM/WMAP5 model? This question is addressed by calculating the angle between \Vbulk\ at a given radius and at $R=10\hmpc$. Figure \ref{fig:align} shows the angle between
${\bf V}{^{\rm (T)}_{\rm bulk}}(10\hmpc)$ and ${\bf V}{^{\rm (T)}_{\rm bulk}}(R)$ for T= WF, CR and RAN as a function of R. Table \ref{table:align} shows some representative numbers for the WF field and the mean and scatter for an ensemble of 20 constrained and random realizations. Table \ref{table:align}  and Figure \ref{fig:align}  show that the bulk velocity vector can zigzag quite drastically in unconstrained realization but nevertheless some finite fraction of the realizations retain their coherent direction, akin to our own local flow field.

\begin{table}
\begin{tabular}{lllll}
\hline
R [$\hmpc$]  &  WF  &  CRs  & RANs    \\
\hline
 50     &       11     &      9.6     $\pm$   6.6    &      39   $\pm$    39   \\
100    &       7.5   &       6.6     $\pm$  5.4     &     53    $\pm$    40    \\
150    &       16    &       13      $\pm$  9.2     &     59   $\pm$     36    \\
\hline
\end{tabular}
\caption{The alignment (in degrees)  of ${\bf V}{^{\rm (T)}_{\rm bulk}}(R)$ with ${\bf V}{^{\rm (T)}_{\rm bulk}}(10\hmpc)$. Second column present the alignment for the WF field. The third and fourth columns show the mean and standard deviation for the ensemble of 20 constrained and random realizations.
}
\label{table:align}
\end{table}

\subsection{CMB dipole}

In the standard model of cosmology the CMB dipole anisotropy is induced by the Local Group velocity with respect to the CMB background. This velocity should be recovered as the limiting case of the bulk velocity of a top-hat sphere of a null radius. This is hampered by two possible effects. One is the incomplete sky coverage of the velocities induced by the Zone of Avoidance and the other is possible non-linear effects that might affect the motion of the Local group. Table \ref{table:Vcmb} presents the LG velocity, as inferred from the CMB dipole anisotropy  \citep[ and references therein]{2008ApJ...676..184T}

The comparison shows that the WF recovers the SGX and SGY components of the CMB dipole velocity remarkably well 
and somewhat less well, at $1.7\sigma$, away from the SGZ component. 
The deficiency of the WF reconstruction in the SGZ direction is expected - it is the direction that lies deep in the ZOA and closest to the Galactic center. The surprising result is how well the WF flow field recovers the other two components, and in particular the SGX component that lies in the Galactic plane.

The misalignment of the LG velocity of an ensemble of 15 CRs with the observed dipole is found here to be $(8 \pm 7)^\circ$. However, limiting the study of the alignment in the SGX-SGY plane, namely the Supergalactic Plane, and thereby excluding the SGZ component a much better alignment is found, of $( 0.4 \pm 5 )^\circ$. 

\begin{table}
\begin{tabular}{lllll}
\hline
 & $V_{\rm CMB}$    & $V_{\rm CMB,x}$  & $V_{\rm CMB,y}$   & $V_{\rm CMB,z}$  \\
\hline
Observed  & 631 $\pm$ 20  &      -381                     &   331                                &   -380                               \\
WF/CRs    & 554  $\pm$ 56  &     -364  $\pm$ 56    &   321     $\pm$     50        &   -264      $\pm$   70        \\
\hline
\end{tabular}
\caption{The CMB dipole velocity (norm and the SGX, SGY and SGZ components all in \kms): A comparison of the velocity evaluated at the position of the LG with the observed one. The WF/CRs denotes the  estimated velocity of the Local Group. The one sigma scatter is the standard deviation taken over 20 CRs.
}
\label{table:Vcmb}
\end{table}

\subsection{Statistical significance}

\begin{figure*}
\includegraphics[width=0.75\textwidth]{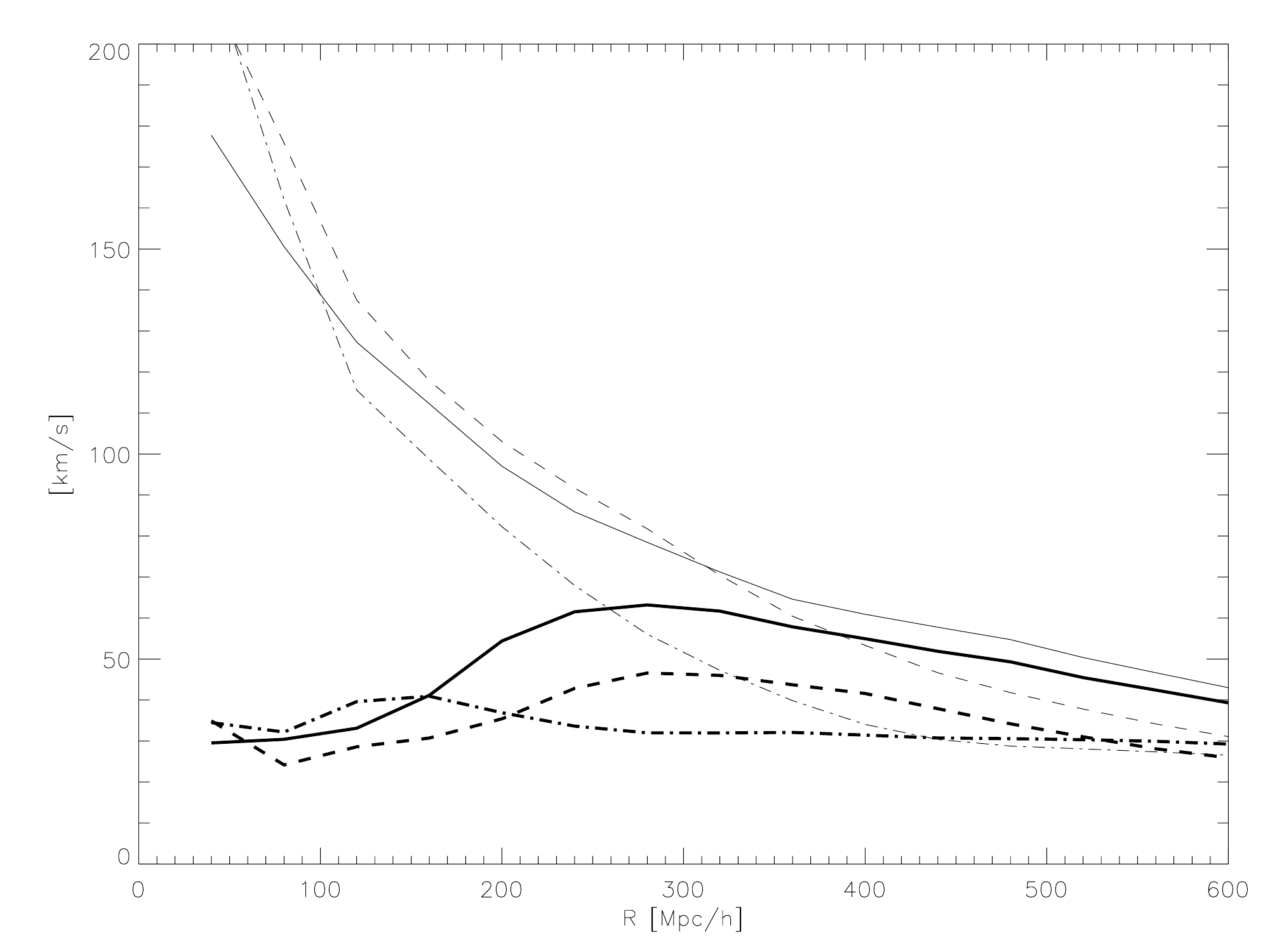}
\caption{The variance (root mean square, rms) around the mean for an ensemble of 20 CRs (thick lines) and of 20 random realizations (RANs).  The plot shows the rms($V_bulk(R)$) for SGX solid line), SGY (dashed) and SGZ (dot-dashed)..
}
\label{fig:CRs-RANs-sigma}
\end{figure*}

The standard deviation of the  Cartesian components of the bulk velocity of the the ensemble of 20 constrained and random realizations is shown in 
Figure \ref{fig:CRs-RANs-sigma}. BOX2560 is used here to calculate the scatter out to $R=600\hmpc$.  The plot shows that out to $R=400\hmpc$ the variance in the value of each of the Cartesian components of the CRs is smaller than cosmic variance. The constrained variance converges to the cosmic one on larger scales, with the rate of convergence varying with the Supergalactic directions.

\subsection{Comparison with SFI++ and the COMPOSITE datasets}
\label{sec:compare}

 \cite{2011ApJ...736...93N} estimated the bulk flow of the local velocity field from the SFI++ survey.
They used ACSE  model (see below), which is close in its approach to the WF/CRs, and their definition of the bulk flow is identical to the one used here. A detailed comparison of \cite{2011ApJ...736...93N} results is shown in Table \ref{table:SFIpp} which shows a close agreement. The CF2 reconstruction has a tighter error bars, by almost a factor of 2, compared with the \cite{2011ApJ...736...93N} numbers.  A discussion of the comparison is presented below, the differences are mainly due to the improvements in the observational data. CF2  is more extended, consists of a larger sample and has a typical 
statistical fractional error  of $\sim 18\%$ compared with the  $\sim 23\%$ of the SFI++.
statistical error of $\sim 23\%$ of distance on individual data-points, while Cosmicflows-2 grouped version (which includes some of the best observational raw data available in SFI++) is somewhat smaller  ($\sim$ 18\%). 

The COMPOSITE compilation \citep{2010MNRAS.407.2328F} is analysed here in the same way as the CF2 data. Namely, the bulk motion of a top-hat sphere of the WF and CRs flow field. Results for $R=50$ and $100\hmpc$ are presented in Table \ref{table:SFIpp}. Again, a good agreement is found with a tighter error bars for the CF2.

\begin{table*}
\begin{tabular}{ccccccc}  
\hline
 R [$\hmpc$] & Data   &   $V_{\rm bulk}$  & Galactic $(l,b)$   & $V_{\rm bulk,x}$  & $V_{\rm bulk,y}$   & $V_{\rm bulk,z}$  \\
\hline
40   &  CF2       & 274 $\pm$ 23 &  $(288^\circ,19^\circ)$     &    -226 $\pm$ 23    &   102  $\pm$ 20   &   -116  $\pm$ 28     \\
       &  SFI++    & 333 $\pm$ 38 &  $(276^\circ,14^\circ)$     &    -242    &   103    &   -203       \\
50   &  CF2       & 250 $\pm$ 21 &  $(280^\circ,18^\circ)$     &    -209 $\pm$ 22    &   90 $\pm$ 19   &   -100  $\pm$ 24     \\ 
 &  COMPOSITE   & 243 $\pm$ 28 &  $(284^\circ,18^\circ)$     &   -193    &   89   &   -1183      \\
100 &  CF2       & 270 $\pm$ 23 &  $(283^\circ,19^\circ)$     &    -211  $\pm$ 31   &   105   $\pm$ 23  &   -131   $\pm$ 33     \\
       &  SFI++    & 257 $\pm$ 44 &  $(279^\circ,10^\circ)$     &    -198    &    61     &   -151       \\
 &  COMPOSITE   & 300 $\pm$ 33 &  $(282^\circ,11^\circ)$     &   -240   &   76   &   -163       \\
\hline
\end{tabular}
\caption{
A comparison of the bulk velocity obtained from the WF/CR reconstruction of the CF2 data and of the  COMPOSITE data compilation and the \citep{2011ApJ...736...93N}
analysis of the SFI++ data. The bulk velocity of a top-hat sphere of radius $R$ is presented in galactic $(l,b)$ and in Supergalactic Cartesian coordinates. Velocities are in units of \kms. 
}
\label{table:SFIpp}
\end{table*}

\subsection{Comparison with Cosmicflows-1  dataset}

Cosmicflows-2 is the second generation catalog of galaxy distances
built by the Cosmicflows collaboration. It has been preceded by the Cosmicflows-1  \citep[CF1][]{2008ApJ...676..184T}  database, which consists of data confined predominantly to redshifts smaller than 3000\kms. It follows that the CF1 is a subset of the more extended CF2 dataset. 
The \citet{2012ApJ...744...43C} study of the CF1 database was  a precursor  of the present paper, using essentially the same methodology at the one used here. Given the shallowness of the CF1 data \citet{2012ApJ...744...43C} focused their efforts mostly on the local cosmography. Appendix  \ref{app-CF1} presents a comparison of the bulk velocity estimated from the CF1 and CF2 data bases. The comparison serves as a manifestation for the power of the WF/CRs Bayesian methodology and it demonstrates how to interpret    the  results of the WF/CRs  analysis. The comparison shows the following expected facts: i.  the  uncertainty in the estimated CF2 bulk velocity is much smaller that those of the corresponding CF1 one; ii. the   CF2 estimated bulk velocity is roughly within the one standard deviation scatter  around the CF1 estimated bulk velocity.

\section{Summary and Discussion}
\label{sec:disc}

The paper reports on the application of the Wiener filter and constrained realizations methodology   to the Cosmicflows-2 data
 set and presents an estimation  of the cosmological bulk flow. A bulk flow is defined here as the mean velocity of a  top-hat sphere centered on the Milky Way.   The bulk velocity has been studied by means of its amplitude, its three Supergalactic  cartesian components and the alignment of its direction across different distances.  
In the limit of vanishing top-hat radius the bulk velocity should converge to the CMB dipole. 
 This condition is indeed the case, with a caveat.  The reconstructed SGX  and SGY components of the LG velocity essentially coincide with the observed values. The SGZ component which lies deep in the ZOA and is most affected by it is roughly two sigma away from the observed value. On a larger  scale, of $R=100\hmpc$, the amplitude of the bulk flow is $270\pm23$ \kms. 
 A non-vanishing signal is discovered all the way to the edge of the reconstruction with $V_{\rm bulk}=59\pm18$ \kms\ at $R=600\hmpc$. The behaviour of the bulk velocity, in terms of direction and amplitude, is consistent with the predictions of  the 
 \LCDM/WMAP cosmological model on all scales, ranging from zero radius out to  $R=600\hmpc$. 
The consistency of the estimated bulk flow with the predictions of the assumed Bayesian prior model constitutes a prerequisite for the present analysis. Indeed, such a consistency has been found here.

The Bayesian framework employed here is standing on two pillars. One is the  WF which provides the mean, and also the most probable, field given the data and the prior model. The other is the ensemble of CRs, which samples the variance of the scatter around the mean field. There is tension between the data and prior model. The WF represents the equilibrium between these two components - the WF solution is dominated by the data where the data is strong and by the prior model where the data is weak. The CRs sample the variance around the WF. The stronger is the data the smaller is the variance of the scatter around the WF field, compared with the unconstrained cosmic variance. It follows that a comparison of the variance of the CRs with the cosmic variance provides a measure of the robustness of the reconstruction. 
The test is best manifested  in Figure \ref{fig:CRs-RANs-sigma}   which presents the variance of the three Supergalactic Cartesian components of the bulk velocity as a function of depth. As expected, the SGZ component is the one  that is least constrained by the data and hence the constrained variance of this component converges to the cosmic variance at $\approx400\hmpc$. For the other two components the inequality extend to beyond $R=600\hmpc$. On the scale of $R=200\hmpc$ the estimated bulk flow is clearly dominated by the data, as is manifested by the fact that for Cartesian components of the bulk flow the cosmic variance is larger than the constrained one by at least a factor of 2.

Two recent studies have focused  on the evaluation of the bulk flow of the local velocity field. \cite{2011ApJ...736...93N} analysed the SFI++ data base by means of the All Space Constrained Estimate (ACSE) model. The model seeks to reconstruct the full space 3D velocity field using a Bayesian regularised fitting of base function sampled from a prior model. Given such a full space reconstruction the bulk flow is determined as the spatial average over a spherical top-hat window. The ACSE model is akin to the current WF/CR method in its use of a Bayesian prior model and definition of the bulk flow, hence its results can be directly compared with the present ones. Table \ref{table:SFIpp} presents a comparison of the bulk flow determined from SFI++ by the ACSE method and  from the CF2 WF/CRs analysis.  There is a good agreement between the two estimates to within one or two standard deviations of each method. Given the different methods of analysis and the different data used such a discrepancy is expected. In both cases the estimated bulk velocity is compatible with the cosmic variance predicted by the WMAP \LCDM\ model.

\cite{2010MNRAS.407.2328F} used a different approach to the problem. They devised an 'optimal minimum variance' weighting scheme and applied it to the COMPOSITE catalog  which consists of a compilation of different peculiar velocities data bases. The Feldman et al method does not provided an estimate of the bulk velocity of a top-hat sphere of a given radius but rather a bulk velocity on an effective scale of an optimally designed window function. It follows that the Feldman et al estimated bulk velocity cannot be directly compared with the ones reported here and with \cite{2011ApJ...736...93N}.  \cite{2010MNRAS.407.2328F} reports a bulk velocity on scales of $\sim 100\hmpc$ of a magnitude of $416 \pm 78$ \kms towards (galactic) $ (l,b)=(282^\circ \pm 11^\circ, 6^\circ \pm 6^\circ)$ 
or (supergalactic) $(sgl,sgb)=(168^\circ , -34^\circ )$. It is further reported that 
``[t]his result is in disagreement with  cold dark matter with  ...  WMAP5  cosmological parameters at a high confidence level''.
The incompatibility of the Feldman et al estimated bulk flow with the WMAP5 \LCDM\ model stands in a sharp disagreement with \cite{2011ApJ...736...93N} and with our results. 
This incompatibility is due to the different mode of analysis of \cite{2010MNRAS.407.2328F} and not due to their data compilation. The application of the WF/CRs methodology to the COMPOSITE dataset results with a bulk flow in full agreement with CF2 and with the  \cite{2011ApJ...736...93N} analysed the SFI++. In particular the WF/CRs bulk flow of the COMPOSITE dataset at $R=100\hmpc$  is  $V_{\rm bulk}=300 \pm 33$ in the direction of (galactic) $(l,b)=(282,11)^\circ$ or (supergalactic) $(sgl,sgb)=(163^\circ , -33^\circ )$.

The standard interpretation of the observed dipole moment of the CMB temperature anisotropies is that it is the outcome of our motion with respect to the CMB. After making a suitable transformation to the LG frame of reference the CMB dipole is being expressed as the LG motion with respect to the CMB, denoted here by ${\bf V}_{\rm CMB}$. The desire to understand the sources of that motion and how it integrates into the large scale velocity field is one of the main motivation for studying the large scale flow field. The current study share this quest. An on-going debate in the community is on the role of non-linear dynamics in shaping the CMB dipole  \citep[e.g. ][]{2011MNRAS.413..585C,2014ApJ...788..157N}. Our analysis sheds some light on the problem. The WF/CRs reconstruction is done within the linear theory framework. Any possible discrepancy between the LG velocity of the WF/CRs flow and the observed one can be attributed to limitations of the method, shortcomings of the data and non-linear effect. 
The present WF/CRs analysis reconstructs the SGX and SGY components of the CMB dipole remarkably well. This is not the case with the SGZ component, with an almost 2 sigma discrepancy between the prediction and observation. The SGZ component lies within the ZOA close to the direction of the Galactic center and thereby it is the Supergalactic component that is most affected by the ZOA. The   good agreement in the SGX and SGY directions and the discrepancy in the direction of the Galactic center suggest a minimal role of  non-linear dynamics and that the discrepancy between the linear theory estimation and observation is dominated by the obscuration within the ZOA.

 The bulk velocity statistics constitutes a powerful statistics and cosmological probe of the large scale structure. The Bayesian methodology of the Wiener filter and constrained realizations has been used here to provide a robust estimation of the bulk velocity of top-hat spheres on scales ranging from the CMB dipole up to roughly 200$\hmpc$. The reconstruction has recovered the observed CMB dipole, away from the direction of the Galactic center, suggesting that the CMB has been hardly affected by non-linear processes.  The resulting bulk velocity is consistent with the  
Bayesian prior \LCDM/WMAP5 standard model up to roughly 200$\hmpc$.
 
\section*{Acknowledgments}   
Fruitful discussions with Jenny Sorce, Daniel Pomar\`ede and Adi Nusser are gratefully acknowledged.
RBT acknowledges support from the US National Science Foundation award AST09-08846 and NAA award NNX12AE70G. 
YH has been partially supported by the Israel Science Foundation (1013/12).
HC acknowledges support from the Lyon Institute of Origins under grant ANR-10-LABX-66 
and from CNRS under PICS-06233.

\appendix

\section{WF, CRs,  $\chi^2$ and all that}
\label{app-wf}

The derivation here presents the essence of the WF/CRs reconstruction. The reader is referred to \cite{1999ApJ...520..413Z}  for a full derivation and for explicit calculation of the covariance matrices. In the following an ensemble of observed peculiar velocities is assumed, $\Big\{ \uo_i \Big\}_{i=1, . . ., N}$ is assumed. It is further assumed that the radial component of the velocity is measured at its true distance, $\br_i$, i.e. the   selection Malmquist bias is nulled.  An individual data point is related to the underlying density field by:
\begin{equation}
\uo_i = \bv(\br_i) \cdot \hat \br_i  + \epsilon_i
\equiv u_i + \epsilon_i .
\label{eq:eps}
\end{equation}

Measurement errors are assumed to be statistically uncorrelated and the errors covariance matrix diagonal and  is written as:
\begin{equation}
 \Bigl <   \epsilon_i  \epsilon_j  \Bigl > = \Big( \sigma{^2_i} + \sigma{^2_\ast}\Big) \delta_{ij}
 \label{eq:err}
\end{equation}
Here $\sigma_i$ is the direct measurement error that typically scales with the distance of the data point. A $\sigma_\ast$  term is introduced to the error covariance matrix as a crude proxy for the contribution of non-linear small scale motions.

The estimated WF velocity field is obtained by a linear convolution on the data vector, $\Big\{ \uo_i \Big\}$:
\begin{equation}
\bv\WF(\br) = \Bigl < \bv(\br) \uo_i \Bigr >  \Bigl < \uo_i \uo_j \Bigr >
^{-1}        \uo_j
\label{eq:WFv}
\end{equation}

Given a random realization of the underlying velocity field a mock data base is constructed , $\tilde{\uo}_i$. This is used to construct a constrained realization:
\begin{equation}
\bv \CR(\br) = \tilde \bv(\br) + \Bigl < \bv(\br) \uo_i \Bigr >
\Bigl < \uo_i \uo_j \Bigr > ^{-1} \bigl( \uo_j -\tilde \uo_j \bigr)
\label{eq:CRv}
\end{equation}

Given a prior model for the underlying velocity field and an errors model, the data auto-covariance matrix is written as:
\begin{equation}
R_{ij} \equiv
\Bigl < \uo_i \uo_j  \Bigr > = \Bigl < u_i u_j \Bigr > + \Bigl
                           < \epsilon_i \epsilon_j \Bigr >
   =  \hat \br_i \Bigl < \bv(\br_i) \bv(\br_j)  \Bigr > \hat \br_j
+ \Big(\sigma{^2_i}+ \sigma{^2_\ast}\Big) \delta_{ij} .
\label{eq:Rij}
\end{equation}

The reduced $\chi^2$ of the data given a prior model is defined by:
\begin{equation}
\chi^2 = {1\over N} \ { \uo_j R{^{-1}_{ij} }\uo_i  \over 2}
\label{eq:chi2}
\end{equation}
The value of $\sigma_\ast$ is tuned to yield  $\chi^2=1.0$.

\section{Convergence test}
\label{app-convergence}

\begin{figure*}
\includegraphics[width=0.49\textwidth]{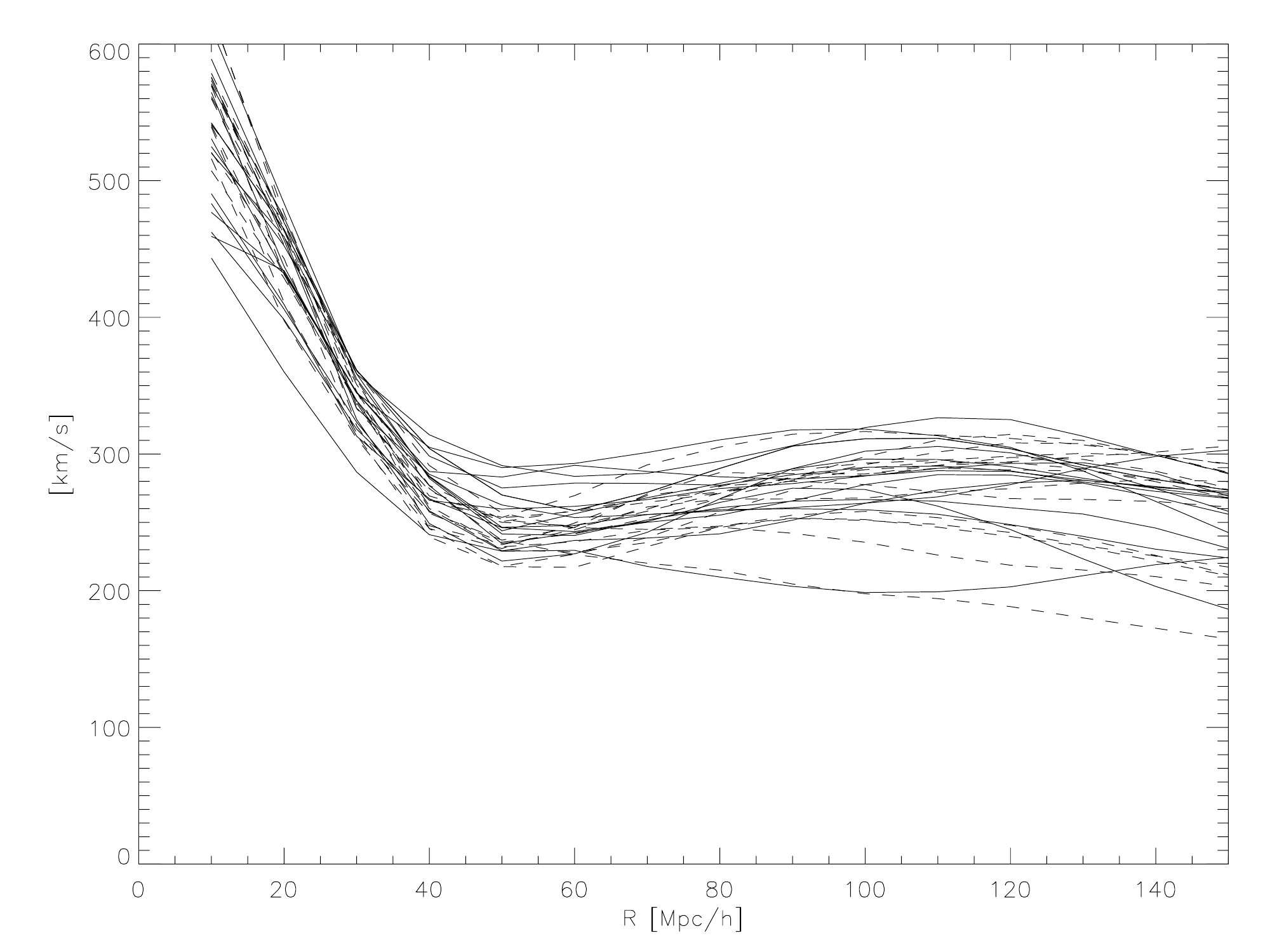}
\includegraphics[width=0.49\textwidth]{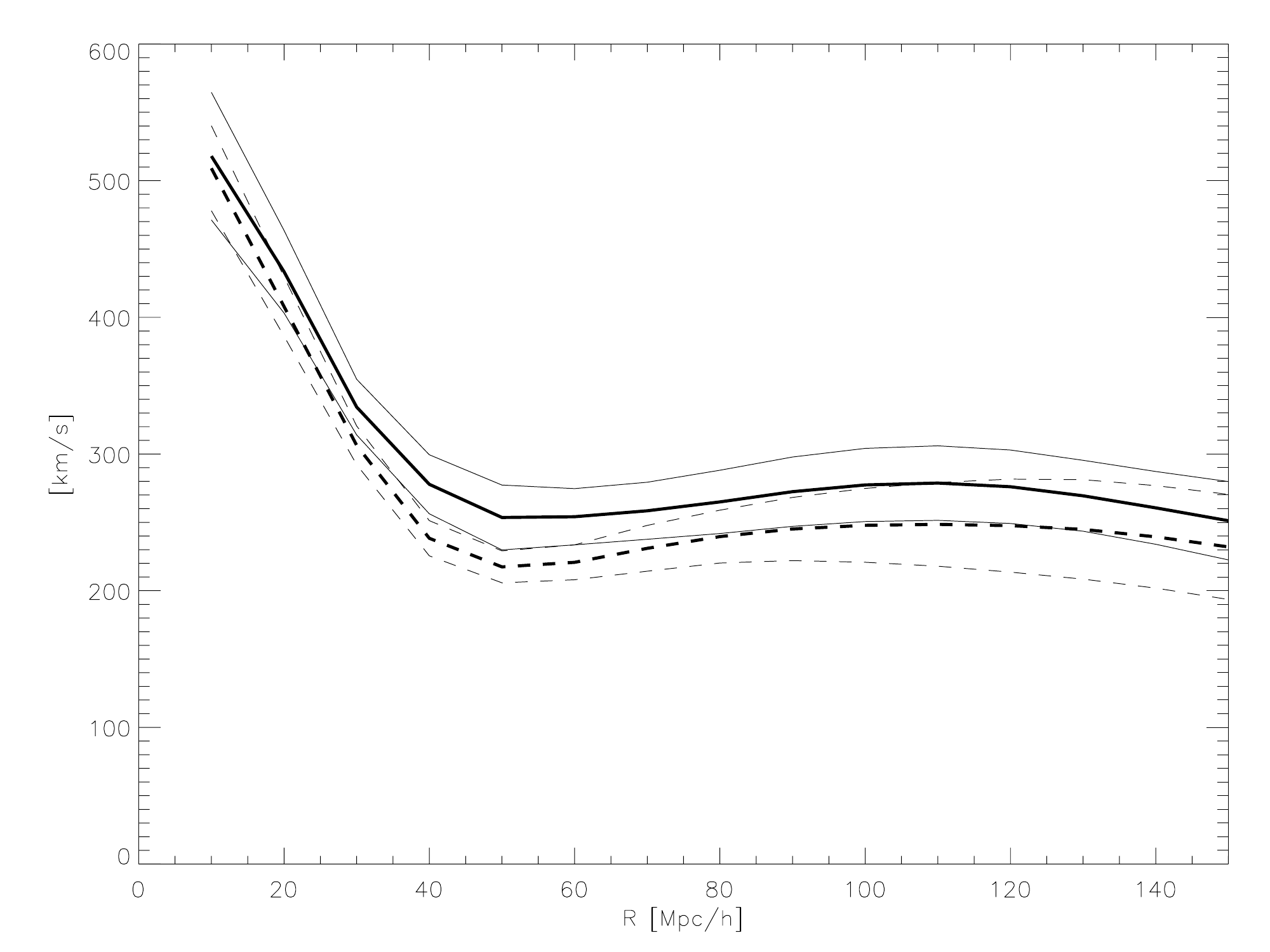}
\caption{
A convergence test on the size of the computational box: Plots show the bulk velocity calculated from an ensemble of 20 CRs calculated within the BOX1280 and BOX2560. Left panel: The bulk velocity of the 20 CRs of the BOX2560 (solid lines) and of the BOX1280 (dashed lines). Right panel: The mean (thick lines) and the mean plus/minus one standard deviation of the distribution of CRs (BOX2560 solid lines, BOX1280 dashed lines).
}
\label{fig:converge}
\end{figure*}

Random realizations of the velocity field are used to construct constrained ones. With the \cite{1991ApJ...380L...5H}   method there is a one to one relation between the random realization (RAN) and the resulting constrained one (CR).  RANs are generated by means of FFT, which imposes periodic boundary conditions. It follows that the RANs do not have any power from outside the computational box. Two boxes are used here, BOX1280 and BOX2560. Figure \ref{fig:converge} presents the amplitude of the bulk velocity of an ensemble 20 CRs constructed within these two boxes. Note that the realizations used in each box are different. The distribution of both boxes are virtually the same and differences are consistent with the scatter expected for an ensemble of 20 realizations.

\section{Bulk velocity and the  MB correction}
\label{app-MBc}

Table \ref{table:MB} presents the bulk velocity, $\VbulkR$, for the WF velocity field calculated from the raw data and the Malmquist corrected data, evaluated at some representative depths. The comparison shows a very small difference in amplitude at $R=10\hmpc$. For $R=50\hmpc$ and larger radii the differences in amplitude amount to a very few \kms, and the differences in the direction are of the order of a very few degrees. The differences between the outcome of the raw and Malmquist bias corrected data are much smaller than the expected scatter around the WF mean field.

\begin{table*}
\begin{tabular}{c c c c c c c c c}
\hline\hline
      & \multicolumn{4}{c}{WF (MB corr.) } & \multicolumn{4}{c}{WF (no MB corr.)}\\
\hline
R        &  & $V_{\rm bulk}$   &    l         &     b     &  &   $V_{\rm bulk}$       &     l      &     b     \\ 
\hline
 10     &  &       536               &   282    &    27    &  &         513                    &   284   &    27   \\ 
  50    &  &       250               &   290    &    18    &  &         249                    &   295   &    18   \\ 
100    &  &       270               &  283     &    19    &  &         274                    &   287   &    19   \\ 
150    &  &       239               &   284    &    11    &  &         239                    &   288   &    13   \\ 
  \hline\hline
\end{tabular}
\caption{
The amplitude and direction (in Galactic ($l,b$) coordinates) of the bulk velocity for the WF velocity field calculated for Malmquist bias corrected data (denoted by 'MB corr'.) and for the raw uncorrected data (denoted by 'no MB corr.'). (Amplitude is given in \kms\ and distances in $\hmpc$.)
}
\label{table:MB}
\end{table*}

\section{Bulk velocity: comparison of the CF1 and CF2 databases}
\label{app-CF1}

The present appendix aims at demonstrating the nature and power of the WF/CRs methodology by considering the estimation of the bulk flow from the Cosmicflows-2 (CF2) and  the Cosmicflows-1 (CF1) databases.  The WF provides a Bayesian estimation of an underlying (velocity) field given the data, its measurement errors and an assumed prior model. The resulting estimation depends on the tension between the quality and strength of the data and the assumed prior model. Where the data is strong  the recovered field is expected  to follow closely the data, otherwise the estimation would tend to the null hypothesis of the prior model. The tell-tale sign of the data dominance is the variance of the constrained realizations around the WF field, which we call here the constrained variance. The constrained variance For  a data dominated constrained field the constrained variance  is smaller  than the cosmic variance, namely the variance of unconstrained realizations. It is the constrained variance which gauges the role of the data and not the value of the WF estimated field.

\begin{figure*}
\includegraphics[width=0.75\textwidth]{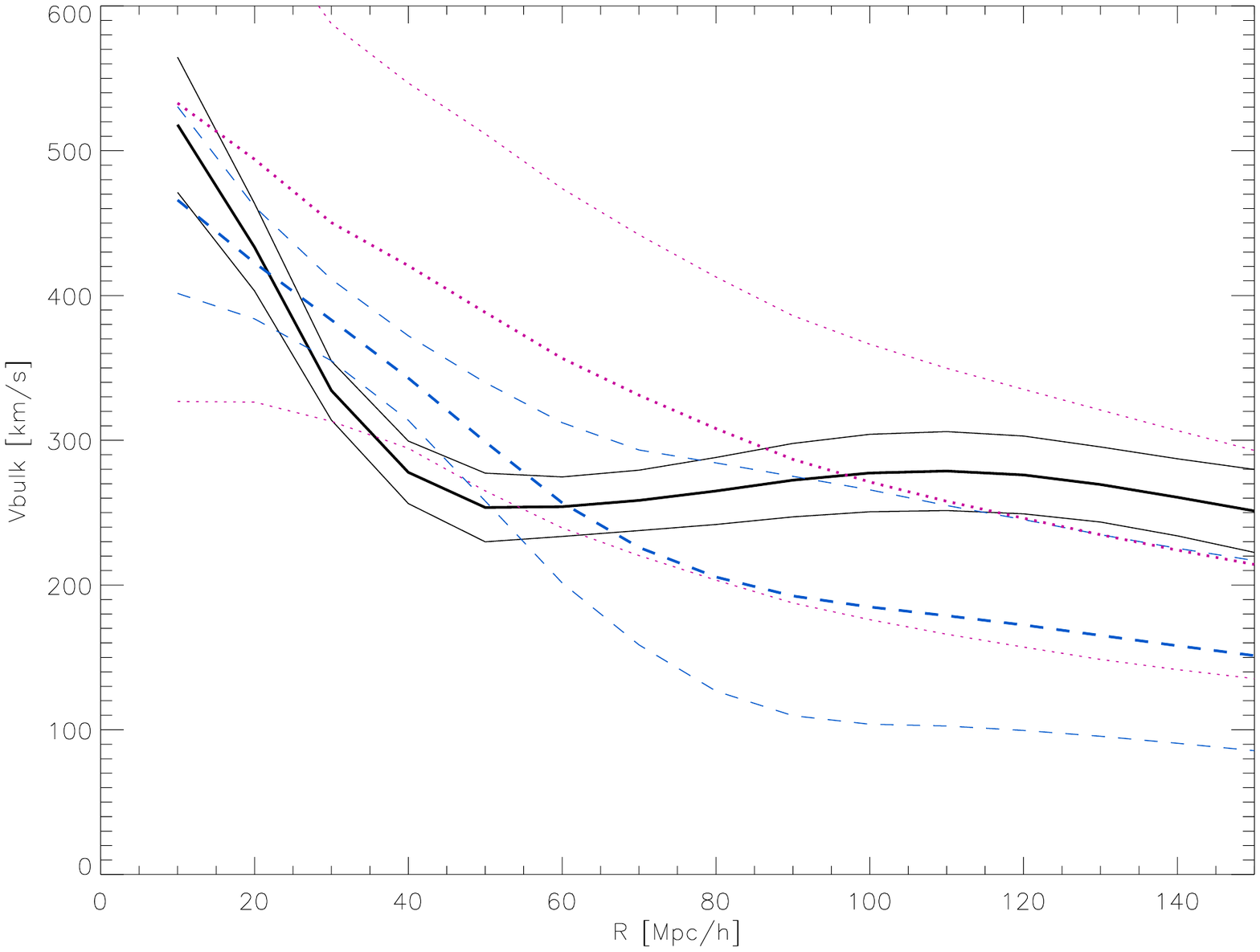}
\caption{Comparison of the mean (WF)  and scatter (CRs)  around the mean of the bulk velocity constructed from the CF2 (solid lines, black) and the CF1 (dashed lines, blue) datasets. The case of unconstrained realizations (dotted lines, red) as well.
The   random realizations used to calculate the mean and cosmic variance are the same ones used to construct the CF1 and CF2 constrained realizations.
}
\label{fig:CR1_CF2}
\end{figure*}

CF2 contains as a subset the  CF1 database, which extends roughly out to 30$\hmpc$. The CF2 is a better database in the sense of its depth and number  of data points.  The comparison of the WF/CRs reconstruction out of the the CF1 and CF2 datasets demonstrates how the quality of the data affects the constrained variance.  Figure \ref{fig:CR1_CF2} shows the WF (mean) bulk velocity as a function of depth, \Vbulk,  and the $\pm$ scatter around the mean of the CRs for the CF1 (in blue) and CF2 (in black) datasets. The case of unconstrained realizations is shown as well (in red). Two main points emerge from the comparison: a. Indeed the CF2 constrained variance is considerably smaller than the CF1 constrained variance and of the cosmic variance (of the random realizations). The CF1 looses its constraining power, in the sense of it constrained variance converging to the cosmic  variance,  at about $R\sim 80\hmpc$. For the CF2 this occurs at $R\sim 300\hmpc$  (Table \ref{table:Vbulk-BOX2560}). b. There is no inconsistency between the WF/CRs reconstruction out of the CF1 and CF2 data. The CF2 mean (WF) \Vbulk\  lies over all scales within the scatter of the CF1 estimated bulk velocity.

\clearpage

\label{lastpage}

\clearpage
\bibliography{biblicomplete}
\end{document}